\documentclass[review]{elsarticle}

\usepackage{hyperref}
\usepackage{amsmath,amssymb,amsfonts}
\usepackage{algorithmic}
\usepackage{ntheorem}
\usepackage{caption}
\usepackage{graphicx}
\usepackage{textcomp}
\usepackage{indentfirst}
\usepackage{float}
\usepackage{bm}
\usepackage{tagging}
\usepackage{amsfonts,amssymb}
\usepackage{amsmath}
\usepackage{tablists}
\usepackage{subfigure}
\usepackage{ragged2e} 
\usepackage{booktabs,makecell, multirow, tabularx}
\usepackage{stfloats}
\usepackage{color}
\usepackage{appendix}
\DeclareMathAlphabet\mathbfcal{OMS}{cmsy}{b}{n}

\journal{Journal of \LaTeX\ Templates}

\begin{document}
\captionsetup[figure]{labelfont={bf},labelformat={default},labelsep=period,name={Fig.}}

\begin{frontmatter}

\title{The Graph feature fusion technique for speaker recognition based on wav2vec2.0 framework \tnoteref{mytitlenote} }
\tnotetext[mytitlenote]{This work has been supported by National Natural Science Foundations of China (No.62071242)}

\author{Zirui~Ge, Haiyan~Guo, Tingting~Wang, Zhen~Yang\corref{mycorrespondingauthor} }
\address{School of Communication and Information Engineering, Nanjing University of Posts and Telecommunications, Nanjing 2100023, China}
\fntext[myfootnote]{E-mail address: yangz@njupt.edu.cn (Z. Yang)}

\cortext[mycorrespondingauthor]{Corresponding author}


\begin{abstract}
Pre-trained wav2vec2.0 model has been proved its effectiveness for speaker recognition. However, current feature processing methods are focusing on classical pooling on the output features of the pre-trained wav2vec2.0 model, such as mean pooling, max pooling etc. That methods take the features as the independent and irrelevant units, ignoring the inter-relationship among all the features, and do not take the features as an overall representation of a speaker. Gated Recurrent Unit (GRU), as a feature fusion method, can also be considered as a complicated pooling technique, mainly focuses on the temporal information, which may show poor performance in some situations that the main information is not on the temporal dimension. In this paper, we investigate the graph neural network (GNN) as a backend processing module based on wav2vec2.0 framework to provide a solution for the mentioned matters. The GNN takes all the output features as the graph signal data and extracts the related graph structure information of features for speaker recognition. Specifically, we first give a simple proof that the GNN feature fusion method can outperform than the mean, max, random pooling methods and so on theoretically. Then, we model the output features of wav2vec2.0 as the vertices of a graph, and construct the graph adjacency matrix by graph attention network (GAT). Finally, we follow the message passing neural network (MPNN) to design our message function, vertex update function and readout function to transform the speaker features into the graph features. The experiments show our performance can provide a relative improvement compared to the baseline methods. Code is available at xxx.
\end{abstract}

\end{frontmatter}

\section{Introduction}
\par Automatic speaker recognition (ASR) is the task of authenticating the claimed identity using the speaker’s voiceprint. As a means of using bio-metrics, ASR has attracted considerable attention from many researchers due to its accessibility and uniqueness. With the development of deep neural network, speaker recognition models based on deep neural networks are being more complicated, and need more larger quantities of labeled training data.
\par However, producing large and high quality labeled data is hard and expensive, and only learning from the labeled samples also seems to be inconsistent with the process of language acquisition of the infants, i.e., self-learning from listening and watching, supervise learning from training and testing with instructors. Self-supervision learning on unlabeled data and fine-tuning on the pre-trained models is similar to the mentioned two stages and have been proved successful for natural language processing such as BERT \cite{Ref1}, GPT-3 \cite{Ref2} \emph{etc}.
\par In the field of speech signal processing, wav2vec2.0 \cite{Ref3} also applies to the two stages learning process. Wav2vec2.0 shows an excellent performance on speech recognition, and it first learns the speech representation from the unlabeled speech audio dataset and fine-tune the pre-trained weights on the labeled data. There are mainly four modules in wav2vec2.0 framework, i.e., a multi-layer convolution feature encoder, a Transformer group module, the quantization module and contrastive loss. More specifically, wav2vec2.0 first encodes the raw audio signal into latent speech representations via the multi-layer convolution feature encoder. Then, the masked latent speech representations are fed into the Transformer module group to capture the contextualized representations from the entire sequence. Meanwhile, the quantization module converts the unmasked latent speech representation into its discrete version via product quantization. Finally, the discrete representation of quantization module and the output of the Transformer group are put into the contrastive loss to identify the true quantized latent speech representation. Wav2vec2.0 framework has achieved 1.8/3.3 word error rate (WER) on the clean/other test sets using all labeled data of Librispeech dataset. When using more less labeled data, wav2vec2.0 still outperforms the state of the art at that time. The excellent performance shows that the phonemic constructions are well learned during the pre-training and the downstream modules can finish their tasks via fine-tuning the pre-training weights. 
\par Pre-trained wav2vec2.0 as the upstream model also works well in speaker recognition. The authors in \cite{Ref4} first applied the wav2vec2.0 to multi-task learning, i.e., speaker recognition and language identification task, and investigated wav2vec2.0 as the audio encoder to extract the speaker and language features \cite{Ref4}. The work of multi-task learning in \cite{Ref4} first demonstrated the effectiveness of wav2vec2.0 on the speaker recognition and language identification task. At the same time, the authors in \cite{Ref5} took pre-trained wav2vec2.0 models to implement the speech emotion recognition task, and propose to weight the output of several layers from the pre-trained model using trainable weights which are learned jointly with the downstream model. The authors in \cite{Ref6} applied wav2vec2.0 framework to speaker recognition task, and investigated effectiveness of different pooling methods. \cite{Ref6} further proposed the first\&cls pooling method that inserts a "start token" (all values are +1) in the input sequence of encoder, and selected the first output token as the speaker embedding. These literatures completed different speech related tasks based on wav2vec2.0 framework via using mean, max, mean\&std pooling methods etc. Though these classical methods have been proven their effectiveness, they only focus on some simple information of features, for example, the mean and mean\&max pooling mainly consider the distribution of features, and max pooling mainly focuses on the "texture information". Besides, these methods consider the output features as many independent elements in the regular (Euclidean) space. They do not view these output features as an entirety feature of a speaker identity and do not consider more complicate relationship among these features. Though GRU as a complicate feature fusion method can focus the temporal information, it may be uncompetitive when meet the features lacking the temporal information. For example, these features have been processed by some other models that can also extract the temporal information, such as the Transformer module \cite{Ref7}. Therefore, when features require to be viewed as an entirety or not merely extracted the temporal information, a new data structure or signal processing technique require to be introduced.
\par Some literatures \cite{Ref8,Ref9,Ref10} have shown that the speech signals can be reformulated as a graph signal and processed using graph signal processing theory (GSP) \cite{Ref25,Ref26} in the irregular space, i.e., graph domain, and a better performance can be obtained compared to regular space. In above literatures, speech features are considered as an entirety, i.e., a graph signal, not merely a set of different independent features. Graph neural networks (GNN) as a nonlinear form of GSP that corporate the advantages of the graph structure and deep neural networks \cite{Ref11,Ref12} have shown the excellent performance on image classification \cite{Ref13,Ref14,Ref15} which is mainly processed in the regular (Euclidean) space previously. Thus, exploring the combination of GNN and speaker recognition is well-founded.
\par The authors in \cite{Ref16} first proposed the concept of GNN and extended the neural networks for processing the data resided in graph domains. With the development of GNN, there are some major variants of GNN in the world including the message passing neural network (MPNN) \cite{Ref17}, graph attention network (GAT) etc. MPNN contains two phases, a message passing phase and a readout phase. The message passing phase mainly include hidden states updating and graph vertex aggregating, and the readout phase is aim to obtain the whole graph representation using the readout function. GAT incorporates the self-attention mechanism \cite{Ref7} into the propagation step and obtains new node features via weighting neighborhood node features using attention coefficients.
\par \cite{Ref19,Ref20} first applied GNN as the backend feature fusion method of ResNet and RawNet2 model to speaker recognition task. \cite{Ref20} uses the GAT mechanism to design an undirected graph with asymmetric weight matrix to depict the relationship between different graph vertex pairs, and take the graph U-Net architecture to obtain the final graph representation. The GNN related framework in [19] takes a pair of utterances as the enrollment and the test utterance to train their similarity score. \cite{Ref21} also took the same GNN architecture in \cite{Ref19} to implement speaker anti-spoofing task. The both GNN backend models obtain the improved performance. However, that mentioned utterance-pair classification format is inferior to the single-utterance classification for speaker recognition based on wav2vec2.0 framework \cite{Ref6}. Besides, both works do not talk about the relationship between the GNN fusion method and classical fusion method such as mean, max, random, etc., and why the GNN based feature fusion models can outperform the compared GRU model.
\par Under this background, we first discuss the relationship between GNN or GSP and classical pooling methods, such as mean, max, random, etc., and show that why GNN based feature fusion models can outperform the mean, max, random pooling methods theoretically. Then we propose GNN as the downstream processing framework to explore the output features of wav2vec2.0 framework in non-Euclidean space. Specifically, we model the set of output features as a graph signal, i.e., each feature is considered as a graph vertex’s feature, and the number of graph vertices is equal to the number of output features. Then we use the graph attention mechanism to construct the symmetric weight matrix to capture the similarity between different vertex pairs. The attention weight matrix is applied to our designed message passing graph neural network to obtain the graph embedding which is also the final speaker embedding. When the downstream processing framework is constructed, we take the wav2vec2.0 as the upstream audio feature extractor and fine-tune its pre-training weights. We implement our experiments on the VoxCeleb datasets \cite{Ref20, Ref21}, and the experiments show that proposed graph pooling method can obtain better performance than the classical feature pooling methods. We also explain why GNN can obtain the top performance compared to other feature fusion methods. To the best of our knowledge, this study is the first to apply a GNN to wav2vec2.0 framework for the speaker recognition task.

\section{Related Work}

\par In this section, we first review the pre-training of the wav2vec 2.0 and how to apply the pre-training to downstream. Then, we introduce graph message passing neural network.
\subsection{Wav2vec 2.0 pretraining and fine-tuning}
\par The main body of the model consists a CNN based feature encoder, a Transformer-based context network, a quantization module and a contrastive loss. The feature encoder consists of 7 blocks and each block contains a temporal convolution with 512 channels with respective kernel sizes of (10, 3, 3, 3, 3, 2, 2) and stride (5, 2, 2, 2, 2, 2, 2) followed by a layer normalization and a GELU activation function \cite{Ref21}. As the Figure.1 left depicted, the CNN feature extractor takes as input raw audio $X$ and and outputs latent speech representations $Z$. Then, the latent speech representations are projected into a new dimension, before fed to the following modules.
\par The context network contains 12 Transformer blocks and a residual 2-layer feed forward network with 3072 and 768 units. The relative positional embeddings instead of fixed positional embeddings are first added to the masked speech representations, before the masked representations are input to the context network. Transformer then contextualizes the masked representations and finally generates context representations $C$. The outputs of 12 Transformer blocks and projected speech representation are considered as hidden features.
\par The quantization module is to discretize the output of the feature encoder to a finite set of speech representations via product quantization, where the product quantization is aim to choosing quantized representations from multiple codebooks and concatenating them. The number of codebooks is equal to 2 and there are 320 elements and respective size is 128 in each codebook. The gumbel softmax function \cite{Ref24} is also used to enable choosing discrete codebook entries in a fully differentiable way.
\par The objective function is the weighted sum of the contrastive loss and diversity loss. The contrastive loss requires to identify the true quantized latent speech representation for a masked time step within a set of distractors. The diversity loss is designed to encourage using the codebook entries equally. 
\par Fig.1 right shows the fine-tuning stage, the models are identical to the training stage, except the quantization modules and extra output layers. We take a speech audio with 48000 samples as the input, and show the output shape of different modules. In the fine-tuning stage, how to select the hidden feature has an important effect on downstream tasks \cite{Ref5}, therefore we consider two approaches to select hidden feature, one way is taking the output of last Transformer block, the other is weighting all the hidden features.

\begin{figure}[t]
	{	\centering
		\includegraphics[width=5.2in]{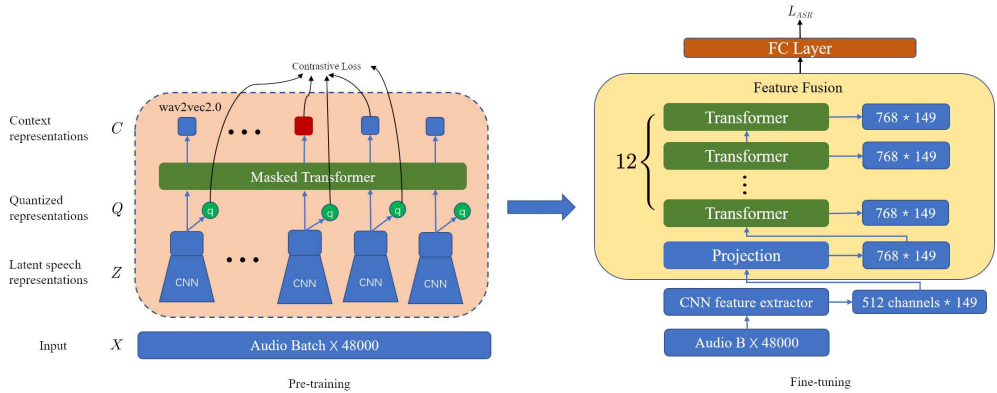}\\}
	\caption{An overview of the pre-training and fine-tuning. The right subfigure shows the output shape of different model for a given input.}\label{fig1}
\end{figure}

\subsection{Message Passing Neural Networks }
\par Let $G=(V,E,A)$ be a graph and $X\in \mathbb{R} ^{F\times N}$ be input vertex features, where $V=\left\{ v_1,...,v_N \right\} $ is the set of $N=|V|$ vertices, $F$ is the dimension of one input feature. $\mathcal{E} =\left\{ e_{i,j} \right\} _{i,j\in \mathcal{V}}$ is the set of edges between vertices, such that $e_{i,j}=1$ if there is a link from node $j$ to node $i$, otherwise $e_{i,j}=0$. $A$ is the $N\times N$ weight adjacency matrix and its entries are the edge weights $a_{i,j}$, for  $i, j=1,...,N$.
\par The MPNN contains two phases, a message passing phase and a readout phase. The message phase runs for $T$ time steps and contains two functions that are message function $M_t$ and vertex update function $U_t$. The hidden state $h_t$ of each vertex at the $t$ time step can be written as:
$$
m_{v}^{t+1}=\sum_{w\in N(v)}{M_t}\left( h_{v}^{t},h_{w}^{t},e_{vw} \right),\eqno(1) \label{eq1}
$$
$$
h_{v}^{t+1}=U_t\left( h_{v}^{t},m_{v}^{t+1} \right) ,\eqno(2) \label{eq2}
$$
where $N(v)$ denotes the neighbors of vertex $v$ in graph $G$. The readout phase computes the final graph feature vector via the following readout function $R$
$$
\hat{y}=R\left( \left\{ h_{v}^{T}\mid v\in G \right\} \right) ,\eqno(3) \label{eq3}
$$
The message function $M_t$, vertex update function $U_t$ and readout function $R$ are all learned differential functions. 

\section{METHODOLOGY}
\label{}
\par In this section, we first show that some classical pooling methods can be represented by GNN or GSP, and introduce the graph neural network as the feature pooling method. The proposed module is located after the context network and takes as input the hidden states of wav2vec2.0. There are three components in our proposed model: 1) the graph attention layer; 2) the message and vertex update function; and 3) the readout function. Specifically, we first show how to obtain the graph weight adjacency matrix from the graph attention layer. Then we show how to aggregate neighbor vertices information in term of one vertex, and update the status of every vertex. Finally, we propose a readout function to obtain the whole graph feature embedding. The overall scheme is illustrated in Figure 2.
\subsection{Reformulating some classical pooling methods with GNN}
\par We reformulate some pooling methods using GSP and GNN. Specifically, we use the GSP to reformulate the linear pooling, including mean, random, first, middle, last pooling method, and use the GNN to reformulate the nonlinear pooling, i.e., the max pooling method.
\par In the graph signal processing theory, the weight adjacency matrix $A$ is also is also named as the shift operator, and the notion of graph shift operator is defined as a local operation that replaces a graph signal feature with the linear combination of features at the neighbors of that vertex \cite{Ref27}. The graph shift operation can be expressed as 
$$
Y=AX^T, \eqno(4) \label{eq4}
$$
where $X\in \mathbb{R} ^{D\times N}$.
\par We take the above graph shift operator to reformulate the mean and random pooling methods. When 
$$
A=\left[\begin{array}{ccc}
1 / N & \cdots & 1 / N \\
\vdots & \ddots & \vdots \\
1 / N & \cdots & 1 / N
\end{array}\right], \eqno(5) \label{eq5}
$$

the expression $A^T$ can be considered as the mean pooling method, and for 

$$
A=\left[\begin{array}{ccc}
0 & 1 & 0 \\
\vdots &\vdots & \vdots \\
0 & 1 & 0
\end{array}\right], \eqno(6) \label{eq6}
$$
the expression $A^T$ can be considered as the random pooling method. When the random pooling method selects the $i$th feature, the entries of the $i$th column in $A$ in are all 1. When the $i$ in random pooling is always 1, $N$, or $\lfloor {{N} /{2}} \rfloor $, the pooling method can be considered as the first, last, and middle pooling.
\par The max pooling method as a nonlinear operation is impossible to be reformulated by the linear operation. We take the MLP as an approximation for the max pooling method, thanks to the universal approximation theorem \cite{Ref28}, and it can be expressed as 
$$
Y=MLP\left( AX^T \right)  , \eqno(7) \label{eq7}
$$
\par Therefore, some classical pooling methods can be reformulated with GSP or GNN, and we may obtain a better performance theoretically, if we directly use GSP or GNN as our pooling methods.
\subsection{Graph attention layer}
\par We first formulate a graph using the output features of wav2vec2.0 framework. Specially, each output feature is considered as a vertex of a graph. Due to these feature does not have an obvious graph structure, we need to specify the edge set and weight adjacency matrix. For this model, we argue that every vertex pair has an edge to link each other, such that the entries $e_{i,j}$ in edge set $\mathcal{E} $ are all 1. Hence set $\mathcal{E} $ means that the entire output features are interacted each other. Next, we show how to obtain the weight adjacency matrix from the graph attention mechanism. Specially, we consider the weights in the adjacency matrix as the degree of similarity between pairs of vertices. There are two main approaches to graph attention mechanism. One approach leverages the explicit attention mechanism to obtain the attention weights such as the cosine similarity between different vertex pairs \cite{Ref29}. The other approach does not rely on any prior information and leverages complete parameter learning to gain attention weight \cite{Ref30}. In this study, we take the first approach as our graph attention layer. Compared to \cite{Ref30} used in \cite{Ref20}, the symmetric weight adjacency matrix learned in \cite{Ref29} can save half the computation in the graph attention layer. The GAT process can be described as:
$$
b_i=Wx_i, \eqno(8) \label{eq8}
$$
$$
a\left( i,j \right) =\frac{\exp \left( \beta \cos \left( b_i,b_j \right) \right)}{\sum\nolimits_{k=1}^N{\exp \left( \beta \cos \left( b_i,b_j \right) \right)}}, \eqno(9) \label{eq9}
$$
where $x_i\in \mathbb{R} ^F$ is the vertex feature, and $W\in \mathbb{R} ^{F'\times F}$ is the projection matrix, $\beta $ is the learnable parameter. In the GAT layer, the vertex feature is first projected into $F'$ dimensional space via multiplying $W$. Then the attention score is obtained by equation (9). 
\subsection{Message and vertex update function}
\par In this study, we formulate the message function as 
$$
M_t=AH_{t-1}^{\top}, \eqno(10) \label{eq10}
$$
where $H_t\in \mathbb{R} ^{F'\times N}$ is the set of the hidden states $h$, $M_t$ is the aggregated states set, and the subscript $t$ is the $t$th massage aggregation, $\top $ is the transpose symbol. (10) can also be considered as the one order graph shift. When $t=1$, $H_{t-1}$ is the projected graph vertex features set $B=\left\{ b_i \right\} ,i=1,...,N$. 
\par The vertex update function is defined as the nonlinear transformation of aggregated representation:
$$
H_t=\sigma \left( LN\left( MLP_t\left( M_t \right) \right) \right) , \eqno(11) \label{eq11}
$$
where $\sigma \left( \cdot \right) $ is the non-linear activation, $LN\left( \cdot \right) $ is the layer normalization function, $MLP\left( \cdot \right) $ is the multi-layer perceptron (MLP) with a set of learnable parameters.
\subsection{Readout function }
When all the hidden states are sufficiently updated, they are aggregated to a graph-level representation for the speaker voice print feature, based on which the final prediction is produced. We define the readout function as:
$$
H^T=MLP_{\theta}\left( H^{T-1} \right) \odot sigmoid\left( MLP_{\varphi}\left( H^{T-1} \right) \right) , \eqno(12) \label{eq12}
$$
$$
h_{\mathcal{G}}=\frac{1}{|\mathcal{V} |}\sum_{i=0}^T{\sum_{v\in \mathcal{V}}{h_{v}^{T}}}+\,\,\mathrm{Maxpooling} \left( H^T \right) , \eqno(13) \label{eq13}
$$
where $\odot $ is the element-wise multiplication. In the equation (13), $\frac{1}{|\mathcal{V} |}\sum_{i=0}^{T-1}{\sum_{v\in \mathcal{V}}{h_{v}^{T}}}$ denotes the residual connection in GNNs. In the residual connection, we extract the distribution information of early representation as the supplement for the final representation. The MPNN framework is shown in Figure 2.

\begin{figure}[t]
	{	\centering
		\includegraphics[width=5.2in]{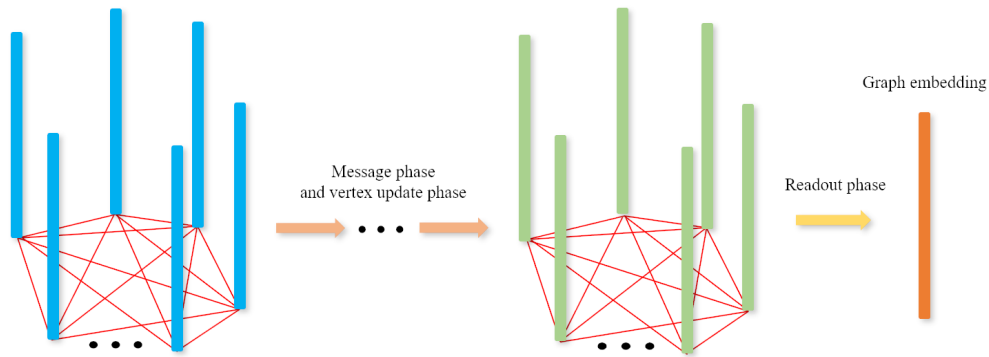}\\}
	\caption{The MPNN framework.}\label{fig2}
\end{figure}

\section{EXPERIMENTS}

\begin{figure}[t]
	{	\centering
		\includegraphics[width=5.2in]{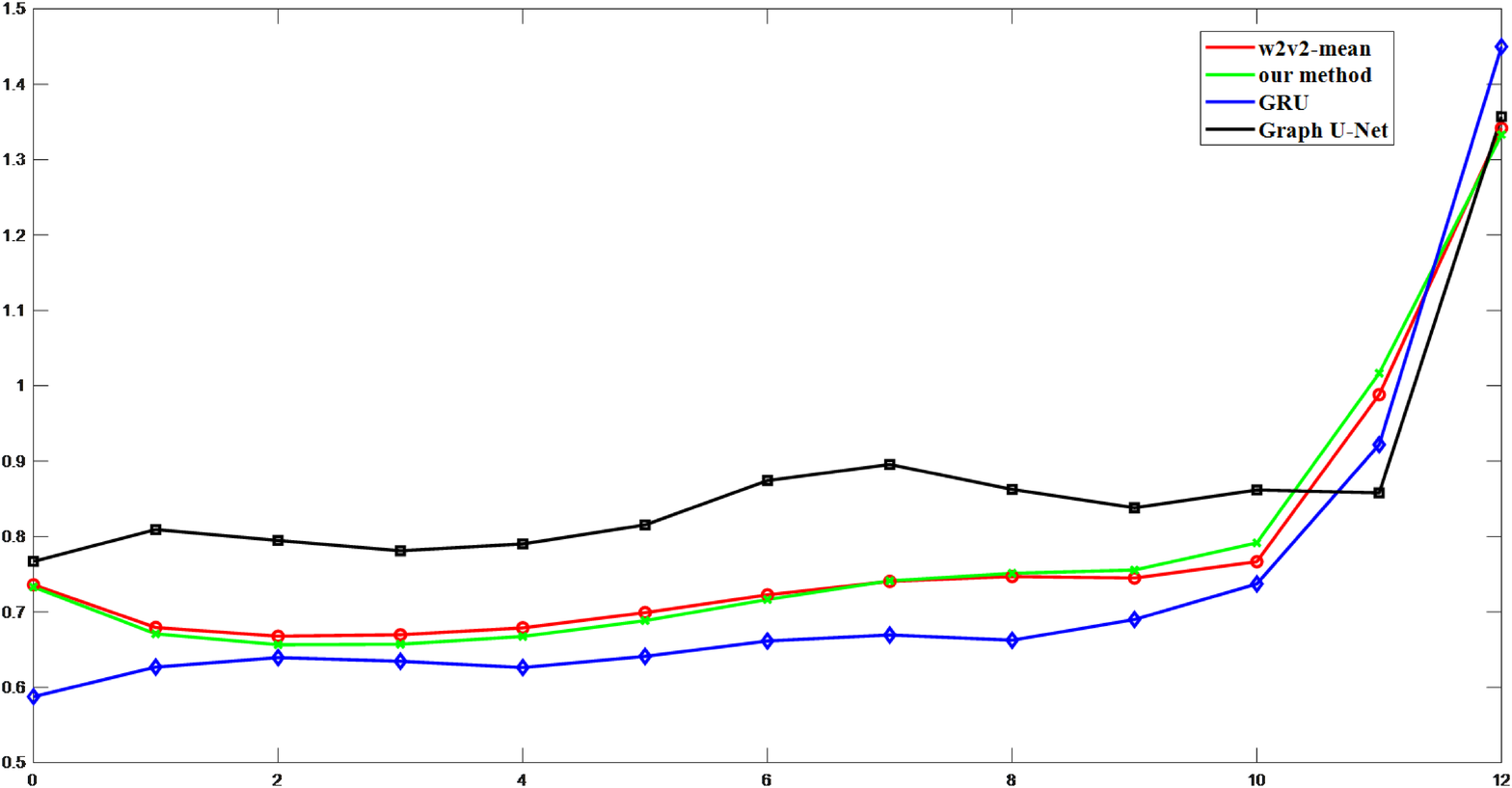}\\}
	\caption{The weights of different hidden features.}\label{fig2}
\end{figure}
\begin{figure}[t]
	{	\centering
		\includegraphics[width=4.3in]{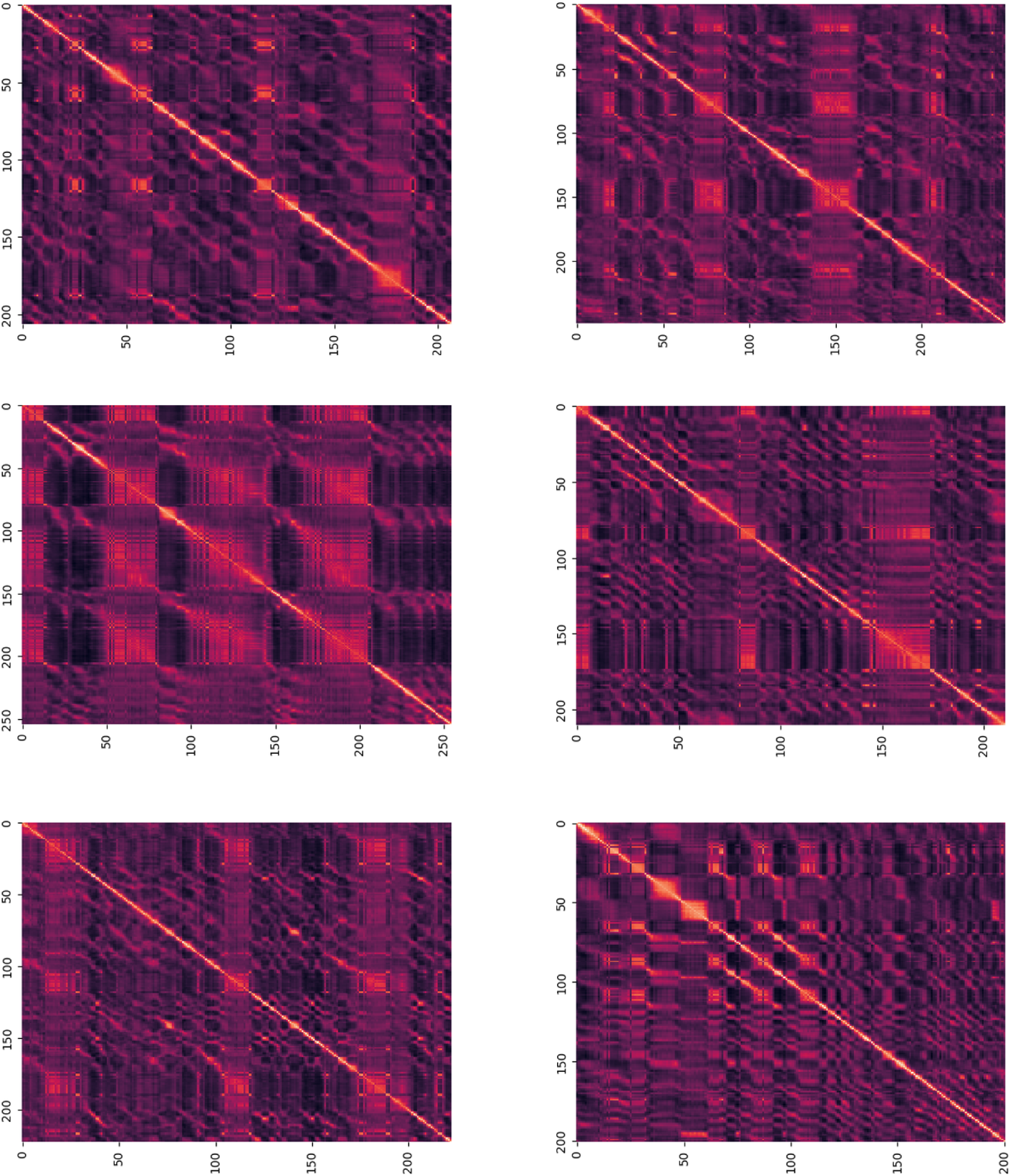}\\}
	\caption{The weight matrices of different speakers.}\label{fig3}
\end{figure}
\subsection{Dataset}
All experiments are conducted on the VoxCeleb1\&2 datasets \cite{Ref22, Ref23}. VoxCeleb2 development set contains over a million utterances from 5994 celebrates from the YouTobe, and the average duration of a signal speaker is about 7.2 seconds. The reported performance in terms of equal error rates (EERs) is evaluated on extended (vox1-o, vox1-e, vox1-h) test sets from the VoxCeleb12. The pretrained weights1 used in the experiments with the wav2vec2.0 framework are released on Hugging-Face \cite{Ref31}.
\subsection{Model description and implementation details}
\par We conduct all experiments using the PyTorch framework, on a 3090 GPU, and take the cosine similarity as the back-end performance evaluation tool.
\par \textbf{Baselines.} In the experiments, we take some released works \cite{Ref4,Ref6,Ref32} for speaker recognition based on wav2vec2.0 framework and feature fusion methods as our baselines. The authors in \cite{Ref6} considered different pooling methods, and they are mean, max, mean\&std, quantile, first\&cls, middle, last, first, random. In this study, we use new symbols to represent them, i.e., w2v2-mean, w2v2-max, w2v2-mean\&std, w2v2-quantile, w2v2-first\&cls, w2v2-middle, w2v2-last, w2v2-first, w2v2-random. The mean pooling is also used in \cite{Ref4} for speaker recognition task, which is restamp as w2v2-mean. Graph U-Net and GRU are also considered as the baseline which are used as a pooling method for speaker recognition in \cite{Ref20,Ref32}. 
\par \textbf{Proposed method.} In the graph attention layer, we set $F'=F$, do not change the dimension of input features, and just project the input features into another space. In the vertex update and readout function, we set $T=2$, the number of hidden layer of ${MLP_t}$, $t=1,2$, $MLP_{\theta}$ and $MLP_{\varphi}$ is 1, and the dimension of hidden layer is 1024 for all the MLPs. 
\par \textbf{Implementation details.} In each experiment, we set batch size is 48, and every sample’s duration is 3 seconds sampling from the audio files. In Table II, we take the pretraining weights of w2v2-mean as the initial weights of wav2vec2.0 for the other models to accelerate the convergence process of the models. We train 30 epochs for w2v2-mean, and train 5 epochs for other models. The optimizer is Adam \cite{Ref33} with a OneCycle learning rate schedule [34], and the loss function is angular additive softmax (AAM) loss function \cite{Ref35,Ref36}. We also propose a thin version of our model, i.e., the   is removed in this version.
\par Similar to the [5], we also weight all the hidden features as:
$$
x=\frac{\sum_{i=1}^{13}{w_i}x_i}{\sum_{i=1}^{13}{w_i}}, \eqno(14) \label{eq14}
$$
where the trainable weights $w_i$ are initialized with 1.0, and the $x_i$ is the output feature of different hidden modules, where $x_1$ s the projected speaker representation, and $x_2$, $\cdots$, $x_{13}$ is the 12 Transformer modules’ output features successively.

\subsection{Results}

\par Table I shows the paraments of different module in each model. From the Table I. we can obtain that our method can reduce the paraments compared to the GRU, but still possesses more paraments than graph U-Net pooling method.
\begin{table}[]
\caption{The number of parameter in different pooling methods.}
\begin{tabular}{llll}
            & wav2vec2 & pooling & loss\_fn \\
our         & 94.4 M   & 6.9 M   & 4.6 M    \\
our/thin    & 94.4 M   & 5.3 M   & 4.6 M    \\
Graph U-Net & 94.4 M   & 3.7 M   & 4.6 M    \\
GRU         & 94.4 M   & 7.9 M   & 6.1 M   
\end{tabular}
\end{table}
\begin{table}[]
\caption{The performance of different pooling methods.}
\begin{tabular}{lllll}

                             &                & voxceleb1 & voxceleb-e & voxceleb-h \\
\multirow{2}{*}{our}         & all\_features  & 1.79      & 1.75       & 3.2        \\
                             & final\_feature & 2.29      & 2.33       & 3.9        \\
\multirow{2}{*}{our/thin}    & all\_features  & 1.83      & 1.88       & 3.22       \\
                             & final\_feature & 1.96      & 1.84       & 3.36       \\
\multirow{2}{*}{Graph U-Net} & all\_features  & 2.07      & 2.2        & 3.99       \\
                             & final\_feature & 1.9       & 1.84       & 3.36       \\
\multirow{2}{*}{GRU}         & all\_features  & 2.1       & 2.07       & 3.92       \\
                             & final\_feature & 2.2       & 2.17       & 4.05       \\
\multirow{2}{*}{max}         & all\_features  & 2.09      & 2.03       & 3.65       \\
                             & final\_feature & 2.04      & 2.03       & 3.65       \\
\multirow{2}{*}{mean}        & all\_features  & 2.14      & 1.9        & 3.89       \\
                             & final\_feature & 1.89      & 1.85       & 3.42      

\end{tabular}
\end{table}

\par Table II shows the performance for different pooling methods based on the wav2vec2.0 framework so that the benefits of proposed methods are assessed. From the Table II, we can obtain that our method can provide a better performance compared to other fusion methods. Our thin model provides a comparable performance and reduce about 23\% parameters compared to the original version. 
\par We also notice that if the only final feature is used as the output feature, the performance of our models and GRU will be degraded, especially in our original model. However, these phenomena are not obvious in other models, and some models even show the opposite results, such as mean, max, graph U-net model. These models all has less parameters than our models and GRU. The number of parameters can determine the upper limit of the expressive power of a model. We believe that the additional information of weighted features beyond the capacity of these models, the extra information may confuse the backend classifier and degrade the performance.
\par Figure 2 shows different weights of each hidden feature, we notice that the final output feature possesses the largest weight compared to other output features in all the models, which is different from the \cite{Ref5}. In the \cite{Ref5}, the larger weights are clustered in the middle features, which means that the emotion features are different from the identity features of a speaker and that is consistent with common sense.
\par It’s worth noting that the GRU based backend classifier does not show the competitive performance compared to other methods. That’s an interesting phenomenon. GRU fusion technique has shown an excellent performance in many models, including speech recognition \cite{Ref37,Ref38,Ref39}, speech emotion recognition \cite{Ref40,Ref41}, speech enhancement \cite{Ref42,Ref43}, speaker recognition \cite{Ref32,Ref44}, etc. We notice that most of these models put the GRU modules right behind convolution modules. That’s reasonable and effective, the convolution modules extract latent speech representation and GRU modules further fused the information in temporal dimension. However, in the wav2vec2.0 framework, the extracted latent representations from the convolution modules have been fed to the Transformer modules, which can also focus the temporal information and has been proved to be excellent at this \cite{Ref7}. Therefore, it is difficult for the GRU modules to extract extra temporal information from the output features of Transformer modules. That is the main reason for GRU possesses the largest number of parameters, but do not show the competitive performance.
\par Mean and max fusion methods mainly focus on the statistical information of output features and the representative features, respectively. These two methods both provide extra information in non-temporal dimension. That is the reason why these two methods can outperform than GRU.
\par GNN uses graph message passing and vertex updating mechanism to obtain the graph structure information in the non-Euclidean space. In Figure 4, each subfigure shows the graph adjacency matrix in different utterances. From the Figure 4, we can obtain that each center feature has high weights with its adjacent features, which means that our GNN can extract the temporal information. Besides, it’s worth noting that the high similarity has no obvious distance property, i.e., the high weights also appear in some features that are far away from the center features. That means that GNN can not only fuse its adjacency features, but also some important features even in a long distance.  Compared to mean, middle, first and random fusion method that allocate equal weight to each feature and the maximum weight to a random feature, our method can treat different features with different weights. In subsection 3.1, we have proved that the mean, max and random fusion method are the special case of GNN and Figure 3 further concretes this experimentally. 

\section{CONCLUSION}
\par In this paper, we mainly take the wav2vec2.0 framework as the speaker feature extractor to apply to speaker recognition task and then investigate the graph neural network as the backend processing tool to aggregate the speaker features. Specifically, we first show that our motivation is reasonable by proving some classical pooling methods can be expressed in the GSP form or GNN form. Then, we obtain the graph structure by GAT and we follow the MPNN framework to design our GNN including message function, vertex update function and readout function. Finally, we evaluate our model on Voxceleb1 dataset, the experiments show the GNN can obtain better performance than the classical pooling method, GRU feature fusion method and the other GNN applied to the speaker recognition task. In experiment results, we show why our proposed method can obtain the best performance compared to other methods.

\bibliography{mybibfile}

\end{document}